\documentclass[twocolumn,showpacs]{revtex4}

\usepackage{graphicx}
\usepackage{dcolumn}
\usepackage{amsmath}
\usepackage{amssymb}

\newcommand{\dotprod}{{\scriptscriptstyle \stackrel{\bullet}{{}}}}

\begin{document}

\title{The stochastic dynamics of nanoscale mechanical oscillators immersed in a viscous fluid}

\author{M. R. Paul}
 \email{mpaul@caltech.edu}
 \homepage{http://www.cmp.caltech.edu/~stchaos}
\author{M. C. Cross}
\affiliation{Department of Physics, California Institute of Technology 114-36,
Pasadena, California 91125}

\date{\today}

\begin{abstract}
The stochastic response of nanoscale oscillators of arbitrary
geometry immersed in a viscous fluid is studied. Using the
fluctuation-dissipation theorem it is shown that deterministic
calculations of the governing fluid and solid equations can be used
in a straightforward manner to directly calculate the stochastic
response that would be measured in experiment. We use this approach
to investigate the fluid coupled motion of single and multiple
cantilevers with experimentally motivated geometries.
\end{abstract}


\maketitle

Single molecule force spectroscopy using nanoscale cantilevers
immersed in fluid is a tantalizing experimental
possibility~\cite{viani:1999,roukes:2000}. The precise manner in
which nanomechanical devices will be utilized for single-molecule
force spectroscopy and sensing  is currently under
development~\cite{arlett:2003,roukes:2000:patent}, however the
detection system will rely upon the change in response of the
cantilever due to the binding of target biomolecules. It is therefore
important to build a baseline understanding of the motion of a fluid
loaded cantilever, or arrays of cantilevers, in the absence of active
molecules. This is the focus of this Letter.

The dynamics of the nanoscale structures considered are dominated by
Brownian fluctuations although the mechanical structures are still
large compared to the molecular size of the fluid molecules. The
elastic response of a single cantilever immersed in fluid has been
investigated previously in the context of atomic force
microscopy~(see for example~\cite{sader:1998,chon:2000}). This
theoretical approach, essentially two-dimensional in nature, modelled
the cantilever as an infinite cylinder oscillating in an unbounded
fluid. This approach has been experimentally validated for long and
slender micron-scale cantilevers~\cite{chon:2000}. However, the
response of nanoscale cantilevers (very strong fluidic damping) or
short and wide cantilevers (where end effects would be important) is
not well understood. Using the fluctuation-dissipation theorem we
show that deterministic numerical simulations of the cantilever-fluid
system can be used to obtain experimentally relevant stochastic
quantities such as the noise spectrum of the fluctuations.

The fluctuation-dissipation theorem allows for the calculation of the
equilibrium fluctuations using standard deterministic numerical
methods. This is possible because the same molecular processes are
responsible for the dissipation and the fluctuations. In the case
under consideration here, predominantly these are the collisions of
the fluid molecules with the cantilever, although dissipative
processes within the elastic material of the cantilever could also be
included.

The fluctuation-dissipation theorem comes in many forms. For the
fluid-damped motion of nanoscale cantilevers it is sufficient to use
a classical formulation. The most convenient form is the one
originally discussed by Callen and Greene~\cite{callen:1952} (see
also~\cite{chandler:1987}). Consider a dynamical variable $A$. For
the classical system of interest here this will be a function of the
microscopic phase space variables consisting of $3N$ coordinates and
conjugate momenta of the cantilever, where $N$ is the number of
particles in the cantilever. We investigate the situation where a
force $f(t)$ that couples to $A$ is imposed.
In this case the Hamiltonian of the system is%
\begin{equation}
H=H_{0}-fA
\end{equation}
and we look at the linear response for very small $f$. Then it can be
shown for the special case of a step function force given by,
\begin{equation}
f(t)=\left\{
\begin{array}
[c]{cc}%
F & \text{for }t<0\\
0 & \text{for }t\ge0
\end{array}
\right. \label{eq:step_force}
\end{equation}
that in the linear response regime, the change in the average
value of a second dynamical quantity $B$ (again any function of
the $3N$ coordinates and
momenta) from its equilibrium value in the absence of $f$ is given by%
\begin{equation}
\Delta\left\langle B(t)\right\rangle =\beta F\left\langle \delta
A(0)\delta
B(t)\right\rangle _{0}%
\end{equation}
where $\beta=(k_{B}T)^{-1}$, $k_B$ is Boltzmann's constant, $T$ is
the absolute temperature, $\delta A=A-\left\langle A\right\rangle
_{0}$, $\delta B=B-\left\langle B\right\rangle _{0}$  and the
subscript zero on the average $\left< \right>$ denotes the
equilibrium average in the absence of the force $f$. Thus we can
calculate a general equilibrium cross-correlation function in terms
of the linear response as
\begin{equation}
\left\langle \delta A(0)\delta B(t)\right\rangle _{0}=k_{B}T\frac
{\Delta\left\langle B(t)\right\rangle }{F}.\label{FDfinal}%
\end{equation}
There are no approximations involved in this result, except that of
assuming classical mechanics and linear behavior. If in addition the
dynamical variables are sufficiently macroscopic that the mean
$\left\langle B(t)\right\rangle $ can be calculated using
deterministic, macroscopic equations, we have our desired result.

There are a number of advantages to a formulation of the
fluctuation-dissipation theorem in time rather than frequency for our
purposes. First, the full correlation function is given by a single
(numerical) calculation, the response to removing a step force. The
spectral properties can be obtained by Fourier transform. This is
particularly advantageous for the low-quality factor ($Q$) situation
characteristic of fluid-damped nanoscale cantilevers, since the
spectral response is very broad, and a large number of
fixed-frequency simulations would be needed to characterize this
response. A second advantage is that no expansion in modes of the
oscillator is needed. Although such an expansion is not too hard for
a high-Q oscillator where the dissipation has negligible influence on
the mode shape, for small cantilevers in a fluid, the coupling to the
fluid is large, and the motion of the fluid complex, so that a mode
analysis would be quite difficult. Finally, the expression
Eq.~(\ref{FDfinal}) allows us to calculate the correlation function
and noise spectrum of \emph{precisely} the quantity measured in
experiment, firstly by tailoring the applied force to couple to one
physical variable measured in the experiment ($A$), and then
determining the effect on the second physical variable ($B$). This
idea has been exploited in the very high-Q situation of the
oscillators used in gravitational wave detectors~\cite{levin:1998},
although there it was convenient to formulate the result in the
frequency domain. In our case for example, if the displacement of the
cantilever is measured through the strain of a piezoresistive layer
near the pivot point~\cite{arlett:2003,roukes:2000:patent} of the
cantilever, it is possible to tailor the force $F$ to couple to this
distortion, and so determine the \textquotedblleft
strain-strain\textquotedblright\ correlation function of one or more
cantilevers.

\begin{figure}[tbh]
\begin{center}
\includegraphics[width=2.5in]{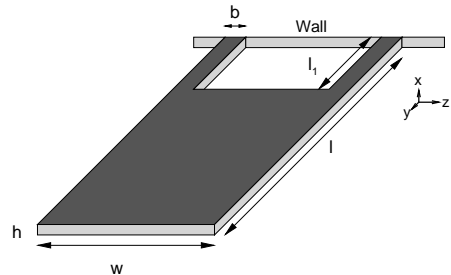}
\end{center}
\caption{Schematic of the cantilever geometry (not drawn to scale):
$l=3\mu$, $w=100$nm, $l_1=0.6\mu$, $b=33$nm. The cantilever is
silicon with a density $\rho_s=2330$Kg/m$^3$, Young's modulus
$E_s=125$GPa. This type of geometry is used to reduce the spring
constant, the equivalent spring constant
$k=8.7$mN/m~\cite{arlett:2003}.} \label{fig:paddle}
\end{figure}

Our scheme consists of the following steps in a deterministic
simulation: (i) apply the appropriate force $f$, constant in time,
small enough so that the response remains linear, and tailored to
couple to the variable of interest $A$, and allow the system to come
to steady state; (ii) turn off the force at a time we label $t=0$;
(iii) measure some dynamical variable $B(t)$ (which might be the same
as $A$ to yield an autocorrelation function, or different) to yield
the correlation function of the equilibrium fluctuations via
Eq.~(\ref{FDfinal}). Using the sophisticated numerical tools
developed for such calculations, we can find accurate results for
realistic experimental geometries that may be quite complex, for
example the triangular cantilever design often used in commercial
atomic force microscopy, or the reduced stiffness geometries
currently under investigation for use as detectors of single
biomolecules as shown if Fig.~\ref{fig:paddle}.

For simplicity, we illustrate our approach by finding the auto- and
cross-correlation functions for the displacements $x_{i}(t)$ of the tips of one or
two nanoscale cantilevers with experimentally realistic geometries. To do this we
calculate the deterministic response of the displacement of each tip, which we call
$X_{i}(t)$ $i=1,2$, after switching off at $t=0$ a small force applied to the tip of
the first cantilever, $F_1$, given by Eq.~(\ref{eq:step_force}). For this case the
equilibrium auto- and cross-correlation functions for the fluctuations $x_1$ and
$x_2$ are precisely
\begin{eqnarray}
\left<x_1(t)x_1(0)\right> &=& k_B T X_1(t)/F_1, \\
\left<x_2(t)x_1(0)\right> &=& k_B T X_2(t)/F_1.
\label{eq:correlations}
\end{eqnarray}
The cosine Fourier transform of the auto and cross-correlation
functions yield the noise spectrum $G_{11}(\nu)$ and $G_{12}(\nu)$,
respectively, which are the experimentally relevant quantities.

We have performed three-dimensional time-dependent numerical
simulations of the deterministic fluid-solid coupled problem
(algorithm described elsewhere~\cite{yang:1994,cfdrc}). The fluid
motion is calculated using the incompressible Navier-Stokes equations
and the dynamics of the solid structures are computed from the
standard equations of elasticity.

For long ($l \gtrsim 100 \mu$) and slender ($l >> w$) cantilevers,
the cantilever response can be well approximated as an infinitely
long oscillating cylinder~\cite{sader:1998,chon:2000}. We first
validate our numerical approach by investigating a cantilever in this
regime. We emphasize that for the experimentally motivated nanoscale
cantilevers of interest here an approximate theory is not available,
yet our numerical approach remains valid providing a means to gain
valuable insight.

\begin{figure}[tbh]
\begin{center}
\includegraphics[width=3.0in]{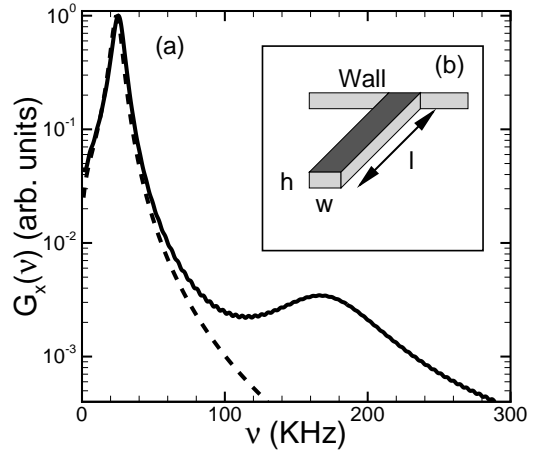}
\end{center}
\caption{Panel~(a) the noise spectrum from simulation (solid line)
and from theory (dashed line). For the theory calculation only the
fundamental mode has been considered. Panel~(b) is a schematic of the
micron-scale cantilever used for validation: length $l=197\mu$, width
$w=29\mu$ and height $h=2\mu$. The applied step force is $F_1=26$
nN.} \label{fig:beam_noise}
\end{figure}

The micron-scale cantilever used for validation has the simple beam
geometry as shown in Fig.~\ref{fig:beam_noise}~(b) (see case c2
in~\cite{chon:2000}). For micron and nano-scale cantilevers immersed
in fluid, dissipation is dominated by the viscous motion of the fluid
driven by the cantilever vibrations. This can be described by a
Reynolds number based on the frequency of oscillation $\omega$ as $R
= \rho \omega w^2/4 \eta$ where $\rho$ is the fluid density and
$\eta$ is the kinematic viscosity. For the cantilevers of interest
here the Reynolds numbers are typically $0.01 \lesssim R \lesssim 1$
indicating that this is in the low Reynolds number regime. Small $R$
corresponds to strong dissipation.

The noise spectrum, $G_{11}(\nu)$, is calculated from the numerical
results by taking the cosine Fourier transform of the
autocorrelations and is shown by the solid line in
Fig.~\ref{fig:beam_noise}(a). The two broad peaks can be identified
with the first two modes of the cantilever. The noise spectrum is
also calculated using the long cylinder analytic approximation and is
shown by the dashed line. The analytical result for the fundamental
mode of the noise spectrum is found in the following
manner~\cite{sader:1998} (note that higher harmonics could be
included if desired). In Fourier space the equation of motion for the
cantilever displacement is
\begin{equation}
 \left( -m \omega^2 + k \right)\hat{x} = \hat{F}_{f} +
 \hat{F}_B,
 \label{eq:eom_fourier}
\end{equation}
where $\hat{F}_f$ is the force felt by the cantilever due to the
fluid,
\begin{equation}
\hat{F}_{f} = m_e \omega^2 \Gamma(\omega) \hat{x},
\label{eq:fluid_force}
\end{equation}
$\hat{F}_B$ is the fluctuating (Brownian) force, $m_e$ is the
effective mass~\footnote{The effective mass is defined such that the
kinetic energy based upon the displacement of the cantilever tip
equals the kinetic energy of the cantilever. For the fundamental mode
of a simple rectangular cantilever (see Fig.~\ref{fig:beam_noise}(b))
$m_e=0.243m$.} of a fluid cylinder of radius $w/2$, $m$ is the
effective mass of the cantilever, and $\Gamma(\omega)$ is the
hydrodynamic function (for an infinitely long cylinder of diameter
$w$ oscillating in the x-direction) which contains both the fluid
damping and fluid loading components and is given
by~\cite{rosenhead:1963},
\begin{equation}
\Gamma(\omega) = 1 + \frac{4i
K_1(-i\sqrt{iR})}{\sqrt{iR}K_0(-i\sqrt{iR})}.
\end{equation}
where $K_1$, $K_0$ are Bessel functions and $i=\sqrt{-1}$. From the
fluctuation-dissipation theorem the spectral density of the
fluctuating force, $G_{F_B}(\nu)$, can be related to the dissipation
due to the fluid and is given by
\begin{equation}
G_{F_B}(\nu) = 4 k_B T m_e T_0 \omega \Gamma_i (\omega),
 \label{eq:brownian}
\end{equation}
where $\omega = 2 \pi \nu$. Solving for the spectral density of the
displacement fluctuations, $G_x(\nu)$, from
Eqs.~(\ref{eq:eom_fourier}),~(\ref{eq:fluid_force})
and~(\ref{eq:brownian}) yields,
\begin{eqnarray}
\lefteqn{G_x(\nu) = \frac{4 k_B T}{k} \frac{1}{\omega_0} \dotprod} \label{eq:xhat} \\
& & \frac{\tilde{\omega} T_0 \Gamma_i(R_0 \tilde{\omega})}{\left[
\left(1 - \tilde{\omega}^2 \left( 1 + T_0 \Gamma_r(R_0
\tilde{\omega}) \right)\right)^2 + \left( \tilde{\omega}^2 T_0
\Gamma_i(R_0 \tilde{\omega}) \right)^2 \right] \nonumber}
\end{eqnarray}
where $\tilde{\omega}=\omega/\omega_o$ is the frequency relative to
the vacuum resonance frequency $\omega_0 = \sqrt{k/m}$, $R_0$ is the
Reynolds number based on $\omega_0$, $T_0$ is the ratio of the mass
of fluid contained in a cylindrical volume of radius $w/2$ to the
mass of the cantilever and $\Gamma_r$ and $\Gamma_i$ are the real and
imaginary parts of $\Gamma$, respectively. Sader's
analysis~\cite{sader:1998} does not take into account the frequency
dependence of the noise force and assumes that the numerator is
constant. The frequency dependence is not large for $R \lesssim 1$;
however the correction has been included in our analysis.
Equation~(\ref{eq:xhat}) yields the analytical curve for
$G_{11}(\nu)$ in Fig.~\ref{fig:beam_noise}(a) and the agreement is
excellent with our numerical results.

\begin{figure}[tbh]
\begin{center}
\includegraphics[width=3.0in]{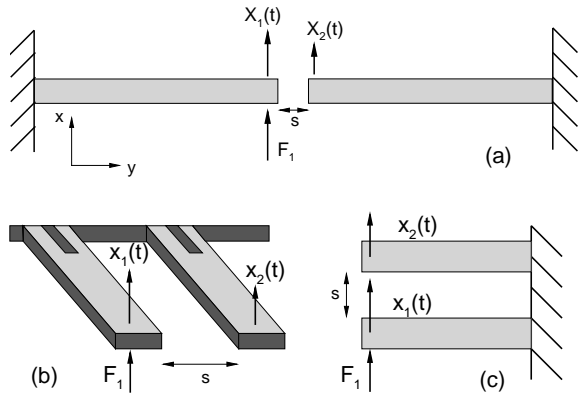}
\end{center}
\caption{Schematic showing various cantilever configurations. In all
configurations the step force $F_1$ is released at $t=0$ resulting in
the cantilever motion referred to by $X_1(t)$. The motion of the
neighboring cantilever is, $X_2(t)$, and is driven through the
response of the fluid. Panel~(a) two cantilevers with ends facing,
panel~(b) side-by-side cantilevers and panel~(c) cantilevers
separated along the direction of the oscillations}
\label{fig:adjacents}
\end{figure}

A recent study of the fluid coupled motion of two adjacent $1\mu$
beads illustrates the importance of understanding the cross
correlations in the fluctuations for use in single molecule force
measurements~\cite{meiners:1999}. The fluid disturbance caused by an
oscillating cantilever is long range, producing motion of the other
cantilevers through the viscous drag. As a result, the stochastic
motion of multiple cantilevers will be correlated. However, the
numerical approach developed here remains valid for multiple
fluid-loaded cantilevers of arbitrary geometry, and our approach can
be used to quantify the response of multiple cantilevers with the
precise complex geometries used in experiment (as we show below) as
well as to help develop a better analytical understanding of
idealized geometries. Various possible cantilever configurations are
shown in Fig.~\ref{fig:adjacents}(a)-(c); here we will present
results for the end to end case and defer results for the other
geometries to a later paper.
\begin{figure}[tbh]
\begin{center}
\includegraphics[width=3.25in]{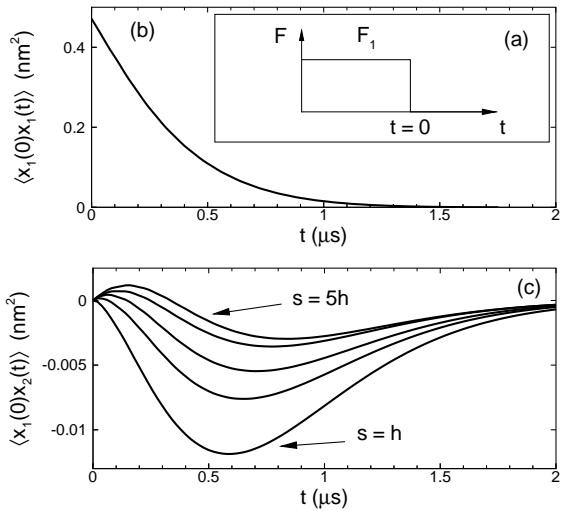}
\end{center}
\caption{Absolute predictions of the auto- and cross-correlation
functions of the equilibrium fluctuations of the cantilevers shown in
Figs.~\ref{fig:paddle} and~\ref{fig:adjacents}. Panel~(a) a schematic
of the step force boundary condition that is applied to the tip of
the first cantilever, $F_1=75$pN. Panel~(b) shows the autocorrelation
and panel~(c) the cross-correlation of the fluctuations (5
separations are shown $s = h,2h,3h,4h,5h$ where only $s=h,5h$ are
labelled and the remaining curves lie between these values in
sequential order).} \label{fig:correlations}
\end{figure}

\begin{figure}[tbh]
\begin{center}
\includegraphics[width=2.0in]{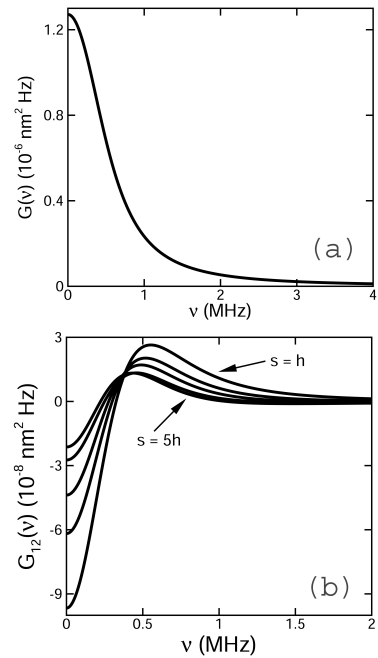}
\end{center}
\caption{Panel~(a): The noise spectrum, $G_{11}(\nu)$. Panel~(b): The
noise spectrum, $G_{12}(\nu)$, as a function of cantilever
separation, $s$, for two adjacent experimentally realistic
cantilevers (see Fig.~\ref{fig:paddle}). Five separations are shown
$s = h,2h,3h,4h,5h$ where only $s=h,5h$ are labelled and the
remaining curves lie between these values in sequential order.}
\label{fig:paddle-noise}
\end{figure}

We use our approach to calculate the behavior of the experimentally
motivated cantilever shown in Fig.~\ref{fig:paddle}. Full
three-dimensional simulations were performed for both one cantilever
and two cantilevers facing end to end in fluid as shown in
Fig.~\ref{fig:adjacents}(a). Through the fluctuation-dissipation
theorem the simulations yield results for the cantilever
autocorrelation function and the two cantilever cross-correlation
function shown in Fig.~\ref{fig:correlations}(b) and~(c),
respectively. The response of the driven cantilever (on the left) was
not significantly affected, for any of the separations investigated,
by the presence of the adjacent passive cantilever on the right. The
value of $\left< x_1(0) x_1(0) \right>$ is $0.471 \text{nm}^2$
indicating that the deflection of the cantilever due to Brownian
motion in an experiment would be $0.686 \text{nm}$ or about $2.3\%$
of the thickness of the cantilever. The cross-correlation of the
Brownian fluctuations of two cantilevers is small compared with the
individual fluctuations. The largest magnitude of the of the
cross-correlation is $-0.012 \text{nm}^2$ for $s=h$ and $-0.0029
\text{nm}^2$ for $s=5h$. The noise spectra for both the one and two
cantilever fluctuations, given by the cosine Fourier transform of the
cross-correlation function, are shown in
Fig.~\ref{fig:paddle-noise}(a) and~(b). Notice that tuning the
separation could be used to reduce the correlated noise in some
chosen frequency band.

The variation in the cross-correlation behavior with cantilever
separation as shown in Figure~\ref{fig:correlations}(c) can be
understood as an inertial effect resulting from the nonzero Reynolds
number of the fluid flow, as follows. The flow around the cantilever
can be separated into a potential component, which is long range and
propagates instantaneously in the incompressible fluid approximation,
and a vorticity containing component that propagates diffusively with
diffusion constant given by the kinematic viscosity $\nu$. For step
forcing, it takes a time $\tau_{v}=s^{2}/\nu$ for the vorticity to
reach distance $s$. For small cantilever separations the viscous
component dominates, for nearly all times, and results in the
anticorrelated response of the adjacent cantilever in agreement
with~\cite{meiners:1999}. However, as $s$ increases the amount of
time where the adjacent cantilever is only subject to the potential
flow field increases resulting in the initial correlated behavior.

In summary, a numerical approach to calculate the stochastic response
of single and multiple nanocantilevers, of arbitrary geometry,
immersed in a viscous fluid using deterministic calculations has been
developed, validated, and applied to an experimentally relevant
cantilever geometry. The methods described here are applicable to
atomic force microscopy in general and also to other nano-structure
fluid interaction problems which are of growing importance as NEMS
technology advances.

This research was carried out within the Caltech BioNEMS effort (M.L.
Roukes, PI), supported by DARPA/MTO Simbiosys under grant
F49620-02-1-0085.  We gratefully acknowledge extensive interactions
with this team.  MP thanks the Burroughs Wellcome Foundation
"Interfaces" program for additional support.

\bibstyle{prsty}
\bibliography{paddle}

\end{document}